\documentclass[aps,prl,twocolumn,showpacs,preprintnumbers,superscriptaddress]{revtex4}

\usepackage{graphicx}
\usepackage{epsf} 
\usepackage{dcolumn}
\usepackage{bm}
\usepackage{amssymb}
\usepackage{amsmath}
\usepackage{amsbsy}

\begin{document}

\title{Self-Organized Ordering of Nanostructures Produced by Ion-Beam Sputtering}

\author{Mario Castro}
\affiliation{Grupo Interdisciplinar de Sistemas Complejos (GISC)
and Grupo de Din\'amica No Lineal (DNL), Escuela T\'ecnica Superior
de Ingenier{\'\i}a (ICAI), Universidad Pontificia Comillas,
E-28015 Madrid, Spain}
\author{Rodolfo Cuerno}
\affiliation{Departamento de Matem\'aticas and Grupo Interdisciplinar de Sistemas 
Complejos (GISC), Universidad Carlos III de Madrid, Avenida de la Universidad 30,
E-28911 Legan\'es, Spain}
\author{Luis V\'azquez}
\affiliation{Instituto de Ciencia de Materiales de Madrid, CSIC, Cantoblanco, 
E-28049 Madrid, Spain}
\author{Ra\'ul Gago}
\affiliation{Centro de Microan\'alisis de Materiales,
Universidad Aut\'onoma de Madrid, Cantoblanco, E-28049 Madrid, Spain}

\date{\today}

\begin{abstract}
We study the self-organized ordering of nanostructures produced by ion-beam
sputtering (IBS) of targets amorphizing under irradiation. By introducing a model
akin to models of pattern formation in aeolian sand dunes, we extend
consistently the current continuum theory of erosion by IBS.
We obtain new non-linear effects responsible for the in-plane ordering of the
structures, whose strength correlates with the degree of ordering found in
experiments. Our results highlight the importance of redeposition and surface
viscous flow to this nanopattern formation process.
\end{abstract}

\maketitle

Performance of many of the (opto)electronic devices currently 
being designed based on arrays of nanostructures such as quantum dots, requires a high 
degree of in-plane ordering \cite{gral_nano}. Currently, there is a formidable
effort to develop experimental techniques which are able to provide highly ordered 
nanostructures in a {\em self-organized} fashion \cite{gral_qdots}. 
These would allow for easy, low-cost and large area fabrication of patterned
structures. Among these techniques, erosion by ion beam sputtering (IBS) at low 
energies \cite{sigmund} 
is a promising candidate \cite{facsko_science,frost,gago}, leading to production of
nanostructures of varying degree of uniformity and order, onto diverse substrates 
such as GaSb, InP, and Si. 
Therefore, detailed knowledge of the basic mechanisms underlying erosion by IBS 
is crucial to understand and control the associated manufacturing process.
From a fundamental point of view, the dynamics of surfaces eroded by
IBS exemplifies neatly the interplay of fluctuations, 
external driving and dynamic instabilities, sharing 
many features with seemingly unrelated non-equilibrium systems, such as 
aeolian sand dunes \cite{sand,dunes_non_cons}. 
Thus, typically surfaces eroded by IBS spontaneously develop submicrometric
patterns (dots, pits, ripples) \cite{reviews_ibs_morph} 
depending on experimental conditions, may deteriorate and eventually 
lead to {\em rough} interfaces, with fluctuations described by the 
universality classes of kinetic roughening \cite{alb}.

A successful approach to surface erosion by IBS is provided 
by continuum evolution equations for 
the surface height, allowing access to time and length scales typical of the 
corresponding pattern formation process.
This approach was pioneered by Bradley and Harper (BH) \cite{bh},
who, based on Sigmund's linear cascade approximation of sputtering in 
amorphous or polycrystalline targets \cite{sigmund}, derived a linear equation that
describes satisfactorily the main features of ripple formation under IBS, such as 
their alignment with the ion beam as a function of incidence angle. 
Additional features, such as ripple stabilization, wavelength dependence with ion
energy or flux, or production of dot or hole structures as a function of bombardment
conditions, required extensions of BH's approach \cite{cuerno_makeev,noisy_ks}, 
leading to a non-linear equation of the Kuramoto-Sivashinsky (KS) type. 
The KS equation provides the continuum description of interfaces appearing in many 
diverse systems, see in \cite{cuerno_makeev},
in which a periodic pattern develops, with a preferred wavelength 
(i.e.\ the lateral size of the nanostructures), that evolves 
into a {\em disordered} array. Thus, the crucial properties of homogeneity and in-plane 
{\em short-range hexagonal ordering} of the nanostructures produced by IBS 
remain to be understood \cite{bobek}. 
Recent attempts have been made at extending the KS equation to overcome 
such shortcomings \cite{facsko2,beyond_ks}, that do not
provide definitive answers, since they either conflict with symmetries of the 
physical system, or with mathematical requirements for well-posedness \cite{comment}.

In this Letter, we present a new continuum model of erosion by IBS. It 
leads to a physically and mathematically well defined generalization of the KS equation
that explains within an unified framework the varying degree of homogeneity and 
order of the nanostructure arrays produced \cite{facsko_science,frost,gago}, as 
a function of experimental parameters.
We exploit connections with 
ripple formation in sand dunes \cite{sand}, hinted at by Aste and 
Valbusa \cite{aste}, overcoming limitations of previous theories 
\cite{bh,cuerno_makeev,noisy_ks,bobek,facsko2,beyond_ks}. 

\paragraph{Model.--}
During IBS, the bombarding ions penetrate the target and induce complex 
collision cascades in the bulk. In semiconductor substrates like
those studied in \cite{facsko_science,gago,frost}, these cascades amorphize 
the near-surface layer. Sputtering events take place when surface atoms
receive enough energy and momentum to break their bonds 
and leave the target. We will assume that only a fraction of those atoms are 
redeposited at the surface. Adatoms are moreover available to relaxation mechanisms 
such as surface diffusion, that can be thermally activated, or else be induced by the 
mentioned change in the local viscosity of the material close to the surface 
\cite{viscousflow}.

In the spirit of the so-called hydrodynamic theory of ripples in aeolian sand 
dunes~\cite{sand}, we define two coupled fields, namely, $R(\mathbf{x},t)$ and 
$h(\mathbf{x},t)$, where $\mathbf{x}=(x,y)$. 
The first one represents the fraction of surface atoms that are {\em not} sputtered
away but, rather, remain mobile along the target surface. Analogously, $h$ measures
the height of the surface neglecting the contribution from the fraction of 
mobile atoms $R$. Time evolutions of $R$ and $h$ are coupled through 
reaction and transport mechanisms \cite{aste}. Thus,
\begin{eqnarray}
\partial_th&=&-\Gamma_{ex}+\Gamma_{ad},\label{heq}\\
\partial_tR&=&(1-\phi)\Gamma_{ex}-\Gamma_{ad}
-{\bf v}\cdot\nabla R-\nabla\cdot {\bf J},\label{Req}
\end{eqnarray}
where $\Gamma_{ex}$ and $\Gamma_{ad}$ are, respectively, the rates of excavation and
addition to the surface, $\mathbf{v}$ is the average velocity of mobile atom, 
and $\phi\neq 0$ 
is the fraction of adatoms that detach irreversibly from the surface. Thus, system 
(\ref{heq})-(\ref{Req}) {\em does not} conserve the amount of material, in marked 
contrast with typical conditions for aeolian sand dunes \cite{dunes_non_cons}.
Here, large redeposition of sputtered atoms corresponds to the 
small $\phi$ limit, while, in the absence of redeposition, $\phi=1$. 
Considering that matter transport along the surface is due to 
diffusion of mobile species, we set $-\nabla\cdot {\bf J}=D\nabla^2R$,
where $D$ is the surface diffusivity.

{\em In the absence of bombardment}, the concentration
of {\em mobile} adatoms $R$ changes due to thermal nucleation of adatoms 
from the ``{\em immobile} state'' $h$, and subsequent transport along the surface. 
Assuming nucleation events are more likely in surface protrusions, we have
$\Gamma_{ad}^{\rm no\,er.}=\tau^{-1}[R-R^0_{eq}(1+\Lambda \kappa)]$,
analogous of the Gibbs-Thompson relation, $\kappa$ being the mean
surface curvature and $\Lambda$ the capillary length, assumed isotropic
due to amorphization by the ion beam. Here $\tau$ is related to the mean time between 
nucleation events, and $R^0_{eq}$ is the mean equilibrium concentration of mobile 
species for a flat surface. {\em In the presence of bombardment}, 
$\Gamma_{ad}^{\rm no\,er.}$ has to be generalized,
to include the contribution of erosion to surface mobility~\cite{viscousflow}.
If the ions fall onto the target along the $x$ direction, forming angle $\theta$ 
with the normal to the uneroded target, we have, for small slopes \cite{tbp,nota4},
\begin{eqnarray}
\Gamma_{ex}&=&\alpha_0[1+\mu_2(\nabla h)^2](1+\boldsymbol{\alpha}_1 \cdot
\nabla h +\alpha_2\nabla^2h) \nonumber \\
&-&\alpha_0[\alpha_3(\nabla h)^2 -\alpha_4 (\partial_x h) (\nabla^2 h)]- 
\beta\nabla^2h , \label{gex}\\ \Gamma_{ad}&=&\gamma_0[R-R_{eq}(1-\gamma_2\nabla^2h)] ,
\label{gad}
\end{eqnarray}
where $R_{eq}$ and $\gamma_i$ generalize parameters in 
$\Gamma_{ad}^{\rm no\,er.}$ so that 
$\gamma_0 = \tau^{-1} + \tau_{ex}^{-1}$, $\gamma_2 = \Lambda + \Lambda_{ex}$,
with $\tau_{ex}^{-1}$ and $\Lambda_{ex}$ being analogs of nucleation time and 
capillary length of erosive origin \cite{sand,viscousflow}. Coefficients 
$\alpha_i \geq 0$ in (\ref{gex}) are related to 
geometric correction factors that take into account the local variation 
of the ion flux with the surface slopes~\cite{tbp}. $E.g.$, for oblique incidence, 
$\alpha_1, \alpha_4 \propto \sin \theta$, and $\alpha_3=1/2$. Likewise, coefficient 
$\mu_2\geq 0$ is related to the {\em local} variation of the sputtering yield with the 
surface slope \cite{new_note}, assumed to have a local minimum for normal incidence, 
while $\beta \geq 0$ measures the efficiency of erosion due to direct impingement of the 
ions onto surface atoms (knock-on sputtering) \cite{reviews_ibs_morph,sigmund}. 
The positive sign of $\alpha_2$ 
implements the {\em physical instability} inherent to Sigmund's theory, by which erosion 
is more efficient at surface depressions than at surface protrusions \cite{sigmund}.
Actually, the analysis presented below will allow us to relate some of these coefficients 
with the parameters characterizing Sigmund's distribution of energy deposition.

\paragraph{Surface dynamics.--}
Our continuum model of IBS, formed by Eqs.\ (\ref{heq})-(\ref{Req}), 
(\ref{gex})-(\ref{gad}), provides a way to introduce systematically all 
relevant physical mechanisms for IBS, differing from that in \cite{aste} in a number
of features. Rather than considering its full solution,
we proceed by deriving an effective equation for the surface height.
As in the experiments of references \cite{facsko_science,frost,gago}, 
we consider the case of ions bombarding the target at normal incidence ($\theta=0$),
thus $\alpha_1=\alpha_4=|\mathbf{v}|=0$ in (\ref{Req}), (\ref{gex}) \cite{nota}. 
After a transient time of order $\gamma_0^{-1}$, Eqs.\ (\ref{heq})-(\ref{Req}) have a 
{\em planar} solution $h_0(t) = -\alpha_0 \phi t$, 
$R_0(t) = R_{eq}+(1-\phi)\alpha_0/\gamma_0$. Perturbing this solution with
periodic waves of the form $h_k = \tilde{h}_k \exp(\omega_k t + {\rm i} 
\mathbf{k} \cdot \mathbf{x})$, and an analogous expression for $R_k$,
amplification/decay of such perturbations is characterized by the dispersion
relation 
$\omega_k=R_{eq}\gamma_0\gamma_2\left(\epsilon\phi k^2-\gamma_0^{-1}(D+\phi A)[1-
\epsilon(1-\phi)] k^4\right)$,
with $\epsilon =A/(R_{eq}\gamma_0\gamma_2)$ and $A=\alpha_0\alpha_2-\beta$. 
If $A>0$ in $\omega_k$, $i.e.$ if sputtering is dominated by collision cascades 
rather than knock-on events, as occurs at low to intermediate energies where Sigmund's
theory is applicable, there is a band of unstable modes that grow exponentially 
fast, with a {\em linear} dispersion relation $\omega_k$ of the expected 
KS type. At this stage, the surface morphology is dominated 
by a periodic pattern whose wave-vector maximizes $\omega_k$. In-plane
isotropy under normal incidence implies dependence of $\omega_k$ on $k=|\mathbf{k}|$ 
rather than the full wave-vector $\mathbf{k}$, thus the surface power spectral density 
is, rather, maximum on a ring \cite{gago,bobek}. Stabilization of this pattern occurs
when its amplitude is large enough that non-linear effects are no longer negligible.
Close to the instability threshold, the rate of erosion is much smaller than the rate of 
addition to the surface. Hence, parameter $\epsilon$ above, which is the ratio between
these two typical rates, is small. We thus can perform a multiple scale expansion by 
introducing time scales $T_1=\epsilon t$ and $T_2=\epsilon^2 t$, and by rescaling length 
scales as $X=\epsilon^{1/2}x$. To lowest non-linear order ${\cal O}(\epsilon)$ and as 
seen in the slow variables ($X$, $T=T_1+T_2$), surface dynamics is described by 
(see \cite{tbp} for details) \cite{nota3}
\begin{eqnarray}
\partial_{T}H & = & - \nu \nabla^2 H - {\cal K} \nabla^4 H + \lambda_1 (\nabla H)^2 - 
\lambda_2 \nabla^2(\nabla H)^2 
\label{final_eq},
\end{eqnarray}
where $H=h_1+ \epsilon h_2$, and 
\begin{eqnarray}
\nu & = & A\phi , \quad {\cal K} = 
\epsilon\gamma_0^{-1}(D+\phi A)[R_{eq}\gamma_0\gamma_2-A(1-\phi)] , \nonumber \\
\lambda_1 & = & \phi\alpha_0(1/2-\mu_2) , 
\label{coeffs} \\ 
\lambda_2 & = & \epsilon \alpha_0(1/2-\mu_2)\gamma_0^{-1}[(D+\phi A)(1-\phi)-
R_{eq}\phi\gamma_0\gamma_2].
\nonumber
\end{eqnarray}
Eq.\ (\ref{final_eq}) with a noise term, has been already employed in
the {\em growth} of amorphous thin films \cite{raible}.
In our context, Eq.\ (\ref{final_eq}) has some 
important limits. {\em First}, in the absence of ion bombardment, $A=\alpha_0=0$, 
$\gamma_0\to \tau^{-1}$ and $\gamma_2\rightarrow \Lambda$, and in the original
variables (\ref{final_eq}) reduces to Mullins' equation for {\em thermal} surface 
diffusion~\cite{Mullins}, $\partial_{t}h= -DR_{eq}\Lambda\nabla^4h$. In the general
case, (\ref{final_eq})-(\ref{coeffs}) include contributions to surface diffusion
that are both thermally activated, and directly induced by the ion beam
as in \cite{viscousflow}. {\em Second}, the BH limit corresponds to $\phi=1$,
{\em i.e.}, no redeposition. While in \cite{aste} the BH limit zeroes out the $k^4$ 
contribution to the analog of $\omega_k$ ---thus making the typical length scale 
of the dot structures remain undefined within linear instability---, here
Eq.\ (\ref{final_eq}) recovers for $\phi=1$ the equation
obtained within BH's approach to Sigmund's theory \cite{cuerno_makeev,beyond_ks}, 
including the fact that the coefficients of the two nonlinear terms have the same signs 
thus making the equation {\em nonlinearly} unstable and mathematically ill-posed 
\cite{beyond_ks,comment}.
Thus, beyond its experimental relevance, redeposition is crucial in order to 
make the theory mathematically sound. On the other hand, the BH limit allows us to 
extract the phenomenological dependence of the parameters in our model with 
characteristics of the collision cascades, such as the ion penetration depth, $a$, 
and the longitudinal and lateral widths $\sigma$, and $\mu$, characterizing the 
Gaussian decay of enery deposition \cite{sigmund}.
Thus, for $\phi\to 1$ we have, in the notation of \cite{cuerno_makeev},
$\alpha_0 = F$, $\alpha_2 = a \mu^2/(2 \sigma^2)$, $\mu_2 = 1 - \mu^2/(2\sigma^2) 
- \mu^2/(2\sigma^4)$, $R_{eq} \gamma_2 = \mu^2/4$, with $F \propto J E /\sigma$,
where $J$ and $E$ are the average ion flux and energy, respectively. 

\begin{figure}[!th]
\begin{center}
\includegraphics[width=0.35\textwidth,clip=]{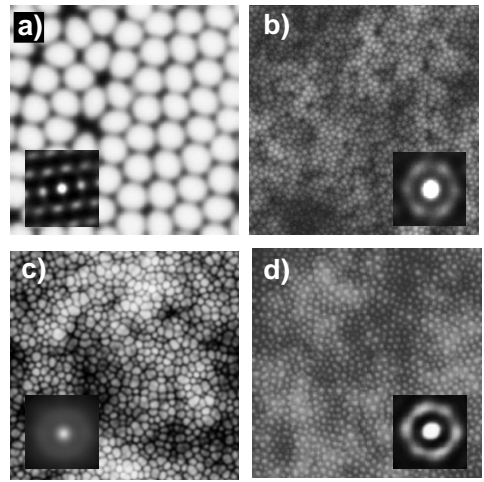}
\end{center}
\caption{$(a)$ Stationary state morphology from the numerical solution 
of Eq.\ (\ref{final_eq}) with $\nu=2{\cal K}=2$, $\lambda_1=0.1$, $\lambda_2=0.5$,
lateral size $L=256$. Units are arbitrary. Inset: 2$d$-autocorrelation, 
showing high degree of short-range hexagonal order. $(b)$ Same as before,
for  $\lambda_1=1$, $L=512$.
$(c)$ same as $(b)$ for the KS equation, 
$\lambda_2=0$. $(d)$ $3\times 3$ $\mu$m$^2$ AFM scan of a Si target
irradiated as in Ref.\ \cite{gago} for 4 h.}
\label{fig1}
\end{figure}

\begin{figure}[!h]
\begin{center}
\includegraphics[width=0.45\textwidth,clip=]{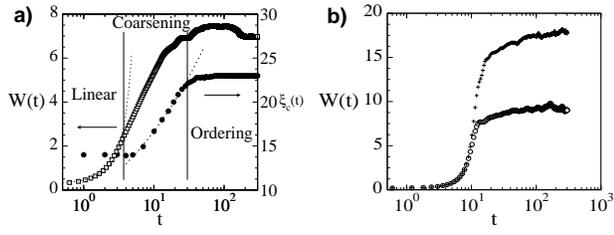}
\end{center}
\caption{$(a)$ $W(t)$ and $\xi_c(t)$ from Eq.\ (\ref{final_eq}) for 
Fig.\ \ref{fig1}$(a)$. Dotted line in the {\em linear} 
({\em coarsening}) region grows as 
an exponential 
(as $t^{0.27}$).
$(b)$ $W(t)$ for Fig.\ \ref{fig1}$(b)$
($\circ$), and Fig.\ \ref{fig1}$(c)$ ($+$).}
\label{fig2}
\end{figure}

Eq.\ (\ref{final_eq}) describes the evolution of the erosion process. Initially,
dynamics is controlled by the linear terms, with the same dispersion relation 
$\omega_k$ as above, and a periodic pattern develops, with characteristic wavelength
given by
\begin{equation}
l_c=2\pi\left[2 R_{eq}\gamma_2 (D +\phi A)[1-\epsilon(1-\phi)]/(A\phi) \right]^{1/2},
\label{length}
\end{equation}
providing the typical size of the nanostructures that form. 
When local slopes become large, the nonlinear terms in Eq.\ (\ref{final_eq}) 
control the dynamics in an opposing way. While the $\lambda_2$ term 
tends to coarsen the nanostructures in amplitude and lateral size, similarly to its 
r\^ole in the coarsening of ripples on aeolian sand dunes \cite{sand}, the nonlinearity
$\lambda_1$ tends to {\em disorder} the pattern leading to the paradigmatic KS 
spatiotemporal chaos. 
{\em Remarkably}, $\lambda_1(\nabla h)^2$ 
seems to {\em interrupt} the coarsening process induced by $-\lambda_2\nabla^2(\nabla h)^2$
and the stationary state morphology consists of domains of hexagonally ordered 
nanostructures separated by defects. The density of these is a function of the 
ratio $r = \lambda_2/\lambda_1$, whose 
$r\to 0$ limit in Eq.\ (\ref{final_eq}) leaves us with the KS equation. 
In Fig.\ \ref{fig1}$(a)$, we plot the stationary-state morphology obtained by 
numerical integration of Eq.\ (\ref{final_eq}) for a relatively {\em large} ratio 
$r=5$ \cite{new_note2}.
The high degree of in-plane short range hexagonal ordering is made clear by the
height autocorrelation function, shown in the inset of Fig.\ \ref{fig1}$(a)$. 
The time evolution of the dot pattern can be assessed in Fig.\ \ref{fig2}$(a)$,
in which the surface roughness (mean height square deviation) $W(t)$ vs $t$ is 
shown 
for the same parameters as in Fig.\ \ref{fig1}$(a)$. 
In excellent agreement with measurements for nanodots on GaSb \cite{bobek}, the roughness
first increases exponentially during development of the linear instability, attains a 
maximum value after dots have coarsened to form a densely packed array, and finally 
relaxes to a smaller stationary value when defects among different dot domains are 
annihilated. Times between linear instability 
and maximum in the roughness correspond to non-linear coarsening of the dot structures,
as seen in the plot 
of the lateral correlation length $\xi_c(t)$, shown on
the same panel. We define $\xi_c(t)$ as the length-scale 
provided by the first secondary maximum of the height autocorrelation.
As seen in Fig.\ \ref{fig2}$(a)$, $\xi_c(t)$scale is constant during linear 
instability, grows as $t^{0.27 \pm 0.02}$, 
and saturates at long times, 
in agreement with experiments on InP \cite{frost}.
This interrupted coarsening process has been also observed on Si \cite{gago}$(b)$
and GaSb \cite{bobek}.

Experimental conditions reflect in the value of $r$ \cite{new_note3}, 
and can be such that this parameter 
is substantially {\em smaller}. Dynamics is then closer to that of the KS 
equation. The intermediate coarsening regime narrows, 
and is followed by {\em kinetic roughening}. A surface morphology produced in these
conditions [$r=0.5$] is shown in Fig.\ \ref{fig1}$(b)$, which
can be compared with an AFM scan [$(d)$] of a Si target irradiated 
as in \cite{gago}. Again, agreement is excellent. 
Note that the morphology now differs appreciably from that of the KS equation,
displayed in Fig.\ \ref{fig1}$(c)$. While for Eq.\ (\ref{final_eq}) a 
{\em short-range ordered} pattern coexists with long-range disorder
and roughening, in the pure KS system {\em disorder} of the cellular
structure is paradigmatic,
see the height autocorrelations in 
Figs.\ \ref{fig1}$(b),(c)$. Still, the time evolution of the roughness
in Fig.\ \ref{fig2}$(b)$ ($\circ$), predicted by Eq.\ (\ref{final_eq})
for small $r$ values, is similar to that of the KS case, 
Fig.\ \ref{fig2}$(b)$ ($+$): initial rapid growth is followed by 
much slower dynamics, and saturation to the stationary state. Such 
is also the experimental behavior found for nanostructures 
produced on Si, see Fig.\ 3 in \cite{gago}$(a)$. Comparing the two plots in
Fig.\ \ref{fig2}$(b)$, for small (non-zero) $r$ values the
small-scale nonlinearity $\lambda_2$ is seen to stabilize the linear instability 
earlier, and leads to smaller stationary roughness.
Moreover, in contrast with Fig.\ \ref{fig2}$(a)$, 
Fig.\ \ref{fig2}$(b)$ shows that for small or zero $r$ values, the roughness 
{\em does not} have a local maximum as a function of time.

In summary, we have introduced a continuum model for the formation of nanometric 
sized patterns by IBS. The model accounts within an unified framework
for experimental features of nanopatterns recently produced on diverse materials.
Moreover, it leads to an effective interface equation providing new predictions. 
Thus, considering dependencies 
\cite{sigmund} on ion energy $E$ of the features of the distribution of
deposited energy, $a$, $\mu$, $\sigma$, the dot size $l_c$ behaves, 
{\em in the large redeposition limit} $\phi \lesssim 1$, as 
$l_c \sim [E + {\rm const.}]^{1/2}$. For small $E$, this implies $l_c$ is energy 
{\em independent}, while $l_c \sim E^{1/2}$ for large enough energies. Observations
exist \cite{bobek,frost2} compatible with such energy dependence, although a systematic
study assessing the importance of redeposition would be highly desirable.

From a fundamental point of view, Eq.\ (\ref{final_eq}) also leads to
new results. Specifically, this is a height equation 
with local interactions in which a pattern is stabilized with constant wavelength and 
amplitude, in contrast with conjectures for 1$d$ systems
\cite{coarsening}. Although more theoretical work is still needed 
[{\em e.g.}, regarding the asymptotic properties of Eq.\ (\ref{final_eq})] this suggests
that in 2$d$ patterns, coarsening dynamics  is 
indeed more  complex than in 1$d$ \cite{interrupted}.

\begin{acknowledgments}
R.\ G.\ acknowledges a Ram\'on y Cajal Fellowship from MECD (Spain). This work has been 
partially supported by MECD (Spain) grants Nos.\ BFM2003-07749-C05, -05 (M.\ C.), 
-01 (R.\ C.), and -02 (L.\ V.).
\end{acknowledgments}

\begin{thebibliography}{99}
\bibitem{nota}
In \cite{frost} $\theta\neq 0$ is used, with simultaneous substrate rotation. 
Results below apply after azimuthal average \cite{tbp}.
\end{thebibliography}
\end{document}